\documentclass[a4paper]{ESASPCS13Style}
\usepackage{epsfig}

\begin{document}

\title{ A Wind analysis of an evolved Giant - {\it{FUSE}} and {\it{HST}}/STIS observations of an eclipsing Symbiotic Binary}


\author{C. Crowley\inst{1} \and B.\,R. Espey\inst{1} \and S.\,R. McCandliss\inst{2}} \institute{Physics Department, Trinity College Dublin , Dublin 2, Ireland
  \and Department of Physics and Astronomy, Johns Hopkins University, Baltimore, MD 21218, USA }

\maketitle 

\begin{abstract}

A major outstanding problem in stellar astrophysics lies in understanding the wind
generation mechanism by which evolved giants lose mass. Phase-resolved observations
of eclipsing symbiotic binary systems, containing a mass-losing giant and a hot white
dwarf, are ideal for studying the wind generation mechanisms in evolved stars. For
such systems we use the orbital motion of the dwarf through the
giant's wind to provide a pencil-beam view through the circumstellar gas. FUV
observations can probe different layers of the wind in absorption, permitting the
derivation of the velocity profile and providing valuable, spatially-resolved
diagnostics of the cool wind.

We present a series of {\it{FUSE}} and {\it{HST}}/STIS observations of two such systems and
discuss our findings. The velocity profiles, and by implication, wind generation
mechanisms for these giants are found to differ from those predicted by commonly used
parametrisations. The phasing of our observations allow us to examine the density,
temperature and velocity structure in the wind acceleration region, as well as the
composition of the outflowing material.

\end{abstract}

\section{Introduction}
  
Stellar winds and the rate of mass lost to the interstellar medium play a central role in stellar and galactic evolution. For hot stars the observed winds are quantitatively explained by line-driven wind generation theory.\ However for cooler, more evolved stars there is no complete theory of mass-loss and understanding the mechanism(s) of mass-loss in late-type giants, in particular, remains a major problem in stellar astrophysics.

For intermediate and low mass stars, mass-loss becomes significant when they approach the end of their lives and enter the red giant (RG) phase (\cite{1992A&AS...96..269S}). Since the wind generation mechanisms are not understood, it follows that the processes that contribute a large fraction of processed material back to the ISM remain unidentified. For cool giants with strong winds it is thought that they lose mass in a two step process. At stellar distances greater than 5 RG radii (R$_{RG}$)  where dust can form, radiation pressure on dust and subsequent dust-gas collisions can explain the observed wind properties (\cite{1994AJ....107.1469D} ). At smaller distances where the wind temperatures are too high to allow dust to form, no convincing models of wind acceleration exist for the case of non-pulsating stars. What is required, therefore, are observational constraints to pin-down the wind conditions and velocity laws in the range 1 - 4 R$_{RG}$ for non-dusty and non-pulsating objects. In general, observations of solitary giants can provide only disk-averaged information on the wind, with only the mass-loss rates and terminal velocities being estimated. However it is possible to probe sightlines through the circumstellar material and spatially examine the wind forming region in some detail by taking advantage of:
\begin{itemize}
\item{the UV spectroscopic capabilities of satellite observatories}

\item{ the rich diagnostic qualities of the  UV bandpass (possible to diagnose the cool gas typically observed in chromospheres as well as much hotter gas seen in coronae)}

\item{and well-timed observations of eclipsing binary systems containing a mass-losing giant and a UV continuum source}
\end{itemize}

\subsection{Symbiotic Stars}

Symbiotic binary systems are the closest separation, non-interacting binaries. They typically consist of a late type giant primary and a hot dwarf or sub-dwarf companion, with line emission originating in a photoionised portion of the RG wind superimposed on the combination spectrum. 

The use of these systems to probe giant winds holds many advantages over other classes of binary stars:
\newline
\hspace*{1. cm} 1) The two components do not directly interact which minimises the effect of interactions on the RG wind itself.
\newline
\hspace*{1. cm} 2) Unlike massive binaries, the two components in symbiotic systems have entirely different spectral characteristics - for example, the hot wind is thin and fast, whereas the cool wind is slow and weakly ionised - permitting the two winds and components to be isolated spectroscopically. 
\newline
\hspace*{1. cm} 3) The WD provides a bright UV continuum source upon which varying absorption from the cool circumstellar gas is superimposed. The FUV and UV is especially rich in resonance and diagnostically informative atomic and molecular transitions which would not be available without the presence of the UV continuum.
\newline
\hspace*{1. cm} 4) These objects have been studied using wavelengths from the radio to x-rays, and the orbital elements and geometrical parameters of many of these stars are well known.

It is therefore possible to use high resolution UV observations of symbiotic stars, taken at well-chosen orbital phases, to tomographically map the structure and conditions in the gas surrounding an evolved giant.

\subsection{\object{EG Andromadae}}

\begin{figure}[ht]
  \begin{center}
    \epsfig{file=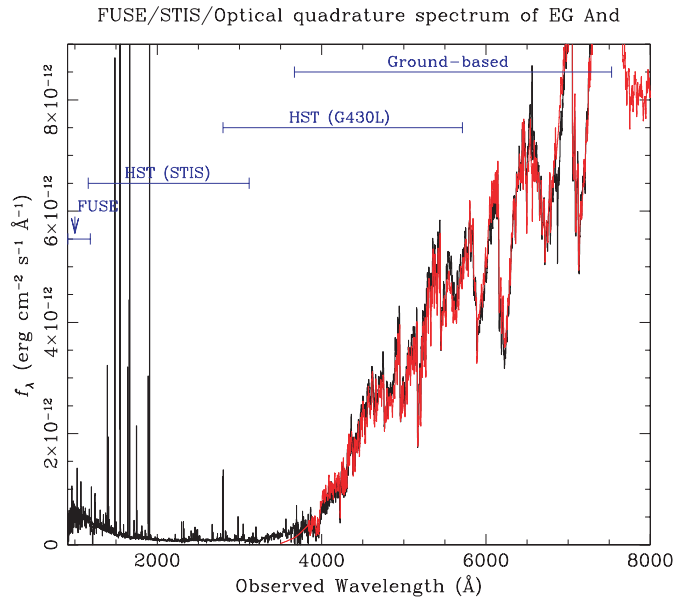, width=8.8cm}
  \end{center}
\caption{A spectrum of EG And, a typical symbiotic system, stretching from the {\it{FUSE}} wavelength region into the optical. The optical region is dominated by the spectrum of the late-type giant.\ The overplotted data (light grey) is an averaged spectrum composed from observations of a number M3 III spectral standards. The close match to the symbiotic giant spectrum suggests that the giant component is relatively undisturbed by the presence of the WD. Note the rising continuum of the WD in the FUV and also the strong emission lines originating in the photoionised region of the wind. \label{fig1}}
\end{figure}

\object{EG And} is the nearest and brightest symbiotic system and consists of a hot, low luminosity  white dwarf  with an M3 giant primary. The UV continuum is known to undergo periodic variations although the system has never been observed to undergo outburst. This variation is attributed to the dwarf being eclipsed by the atmosphere of the primary component and has been observed extensively over several orbital epochs by {\it{IUE}}. The low luminosity of the dwarf means that the giant's atmosphere is less affected by the ionising radiation of its companion than in some similar systems. Shown in Figure\ \ref{fig1} is an uneclipsed spectrum of \object{EG And} combining {\it{FUSE}}, {\it{HST}}/STIS, {\it{HST}} G430L and ground-based echelle data. The binary nature of the object is apparent from the two distinct continuum sources - the WD dominates the ultraviolet and the giant dominates the optical region. The averaged spectra of a number of M3 giant spectral standards is overplotted and is almost identical to the optical spectrum of EG And, graphically showing the similarity of the giant in the binary to isolated giants, despite the presence of its companion.

\begin{table}[bht]
  \caption{System parameters for EG And}
  \label{tab:table}
  \begin{center}
    \leavevmode
    \footnotesize
    \begin{tabular}[h]{lr}
      \hline \\[-5pt]
      \object{EG And} Parameters      \\[+5pt]
      \hline \\[-5pt]
    RG temperature & $\sim$ 3,700 K  \\
    WD temperature & $\sim$ 75,000 K \\
RG Luminosity  & $\sim$ 950 $_{\sun}$\\
WD Luminosity  & $\sim$ 30 $_{\sun}$\\
Period & $\sim$ 481 days\\
Separation & $\sim$ 4.5 R$_{RG}$\\
RG Radius & $\sim$ 75$_{\sun}$\\
Inclination  & $>$ 70$^0$\\
V$_{mag}$ & $\sim$ 7.3 \\
      \hline \\
      \end{tabular}
  \end{center}
\end{table}

The object is well documented in the literature (\cite{2000AJ....119.1375F}, \cite{1993A&A...274L..21V}, \cite{1992A&A...260..156V}, \cite{1984ApJ...281L..75S}) and is considered a prototype for stable, non-dusty symbiotics. The attenuation of the UV continuum (due to Rayleigh scattering by neutral scattering hydrogen in the atmosphere of the giant) around eclipse is readily apparent from the {\it{IUE}} observations. However, the low resolution and signal to noise of the data preclude a more in-depth analysis such as a radial velocity and phase dependent absorption analysis. The higher resolution of {\it{FUSE}} and the E140M setting on STIS, combined with the brightness and low extinction (E$_{B-V}$ $\sim$  0.05) of the target, make it an ideal object to study the wind acceleration region of the giant.

\subsection{Observations}

In total we have obtained 13 {\it{FUSE}} (Far Ultraviolet Spectroscopic Explorer) and 7 Hubble Space Telescope ({\it{HST}}) observations of EG And. The {\it{FUSE}} observations were taken through the large aperture (LWRS) in TTAG photon collecting mode over a time period of three {\it{FUSE}} cycles. The {\it{FUSE}}  satellite covers the wavelength wavelength region 905-1187\AA\ and has been described by \cite{2000ApJ...538L...1M} and \cite{2000SPIE.4139..131S}. See also '{\it{FUSE}} Spectroscopy of Cool Stars' (Harper) in these proceedings.

The 7 {\it{HST}} observations were carried out with the medium resolution ($\lambda/\Delta\lambda\sim$ 30,000 - 45,000) echelle gratings of the Space Telescope Imaging Spectrograph (STIS) through the $0.2^{``}$x$0.06^{``}$ aperture at the 1425, 1978 and 2707 settings.

This therefore provides extensive wavelength coverage stretching from 905\AA\ - 3100\AA . This bandpass encompasses the transitions of many important atomic and molecular species including neutral and low-ionisation states of H, C, N, O, Si, P, Fe, Ni, Mg, Mn as well as CO and H$_2$O. Transitions probing high temperature gas such as OVI, PV, CIV, SiIV, N V, SIV and HeII are also present along with a range of forbidden and semi-forbidden lines.

\begin{figure}[ht]
  \begin{center}
    \epsfig{file=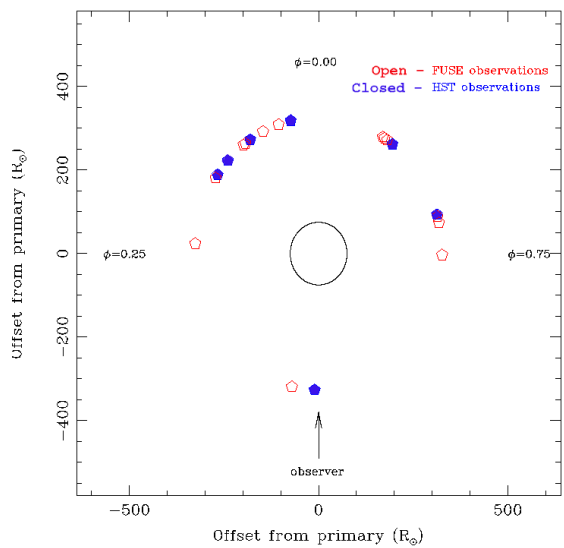, width=8.cm}
  \end{center}
\caption{View of \object{EG And} perpendicular to the orbital plane. Points correspond to the dwarf position for the {\it{FUSE}} and {\it{HST}} observations. Observer's view is from the bottom and the scale is in units of solar radii. \label{fig2}}
\end{figure}

The observing programmes were designed to cover orbital phases both in and out of eclipse. The uneclipsed spectra only show absorption that is either interstellar or else originating in the hot gas associated with the dwarf. These reference spectra can then be compared to observations taken at various degrees of eclipse in order to map and diagnose the circumstellar gas. See Figure\ \ref{fig2} and appendix A for a diagram and log of the ultraviolet observations respectively.

\section{The Wind in Absorption}

\begin{figure*}[ht]
  \begin{center}
    \epsfig{file=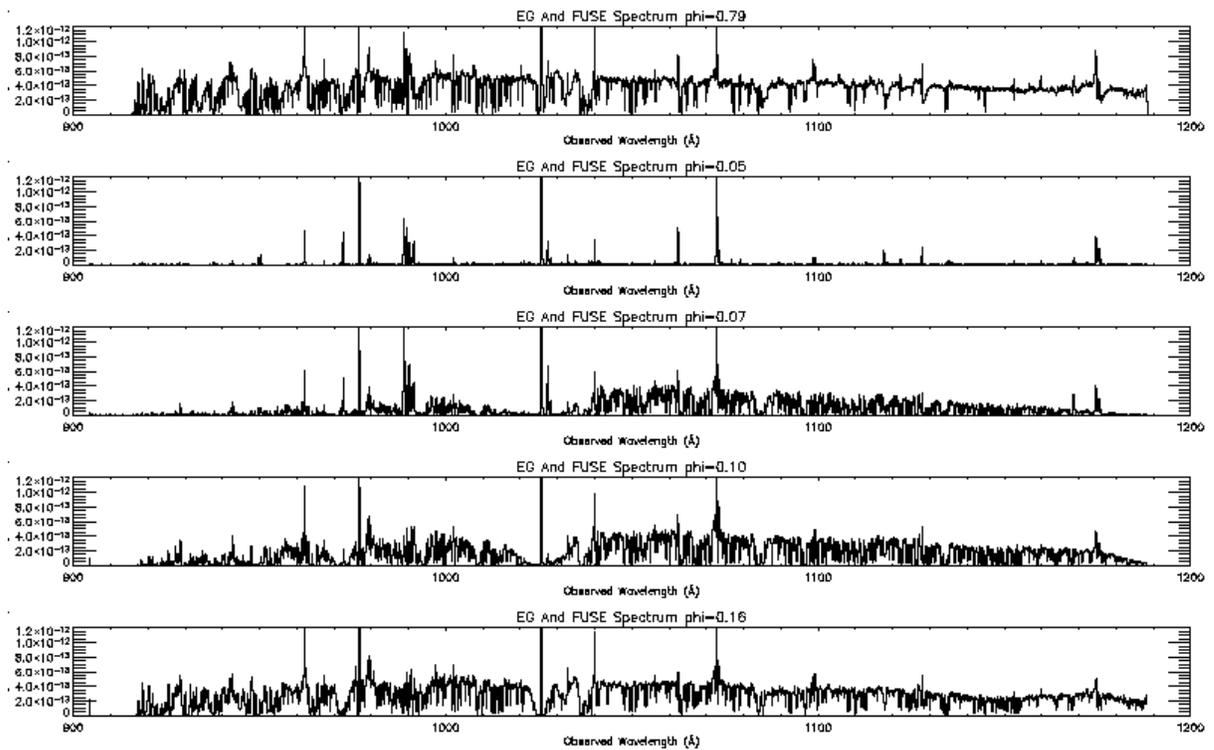 , width=16.cm}
  \end{center}
\caption{ Series of {\it{FUSE}} spectra of EG And going from quadrature, into eclipse, and back out again (top to bottom). Note the varying continuum shapes due to Rayleigh scattering by neutral hydrogen. Its is also possible to see the narrow absorption lines originating in the wind (mostly FeII, NiII and PII in the {\it{FUSE}} bandpass). \label{fig3}}
\end{figure*}

\begin{figure*}[!ht]
  \begin{center}
    \epsfig{file=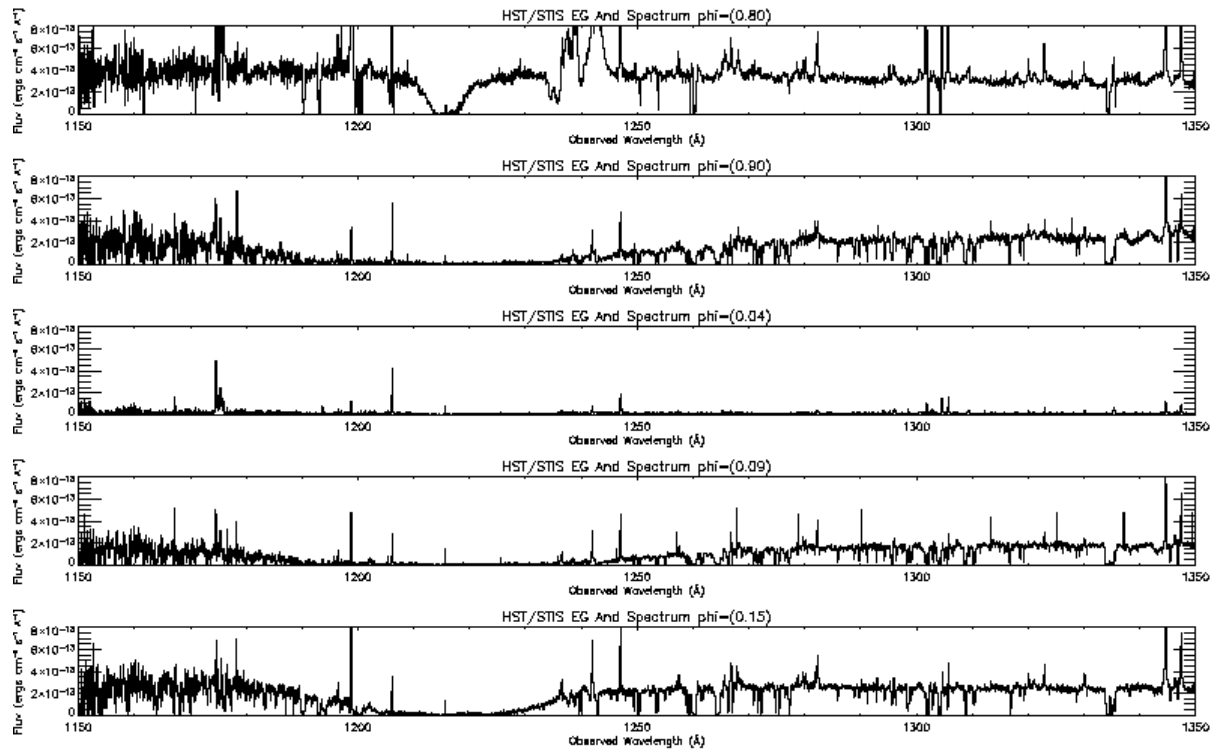 , width=16.cm}
  \end{center}
\caption{ A similar series of STIS spectra. Note how well defined the HI Lyman $\alpha$ profiles are.  \label{fig4}}
\end{figure*}

The spectra are observed to undergo periodic variations with orbital phase due to Rayleigh scattering of ultraviolet photons out of the line of sight during the eclipse of the dwarf. The shape of the ultraviolet continuum during eclipse is defined by the broad damping wings of the hydrogen Lyman series. Superimposed on this continuum during ingress and egress are a host of narrow absorption lines from a range of different low-ionisation species present in the wind. See some sample {\it{FUSE}} and STIS spectra in Figure\ \ref{fig3} and Figure\ \ref{fig4} for an example of the spectral variations around eclipse. The small diameter of the dwarf relative to the giant provides a pencil-beam view through the circumstellar material, while the systemic radial velocity (-95 km s$^{-1}$) means that the wind absorption can easily be distinguished from interstellar absorption. Identifying wind features and continuum placement is also facilitated by comparison with unabsorbed spectra taken out of eclipse. 

\subsection{Low ionisation atomic species}

Identified absorption features in the wind include HI, CI, CII, NI, NII, OI, OII, MgII, AlII, SiII, PII, ArI, MnII, FeII and NiII. The damped profile of H Lyman $\alpha$ provides an accurate measure of the neutral hydrogen column in the line. We use a profile fitting technique to fit the damped wings of Lyman $\alpha$ $\lambda$1216 \AA\ in the STIS spectra, and of  Lyman $\beta$ $\lambda$ 1025 \AA\ and  the blue wing of Lyman $\alpha$ $\lambda$1216 \AA\ in the {\it{FUSE}} spectra.

To determine the column densities of other species whose lines are not on the damped region of the curve of growth we use a combination of profile fitting and the apparent optical depth technique (\cite{1991ApJ...379..245S}). Although there are problems involving line saturation, blending and atomic data uncertainties in the ultraviolet it is possible to constrain the columns for many species quite accurately.

\begin{figure*}[ht]
  \begin{center}
    \epsfig{file=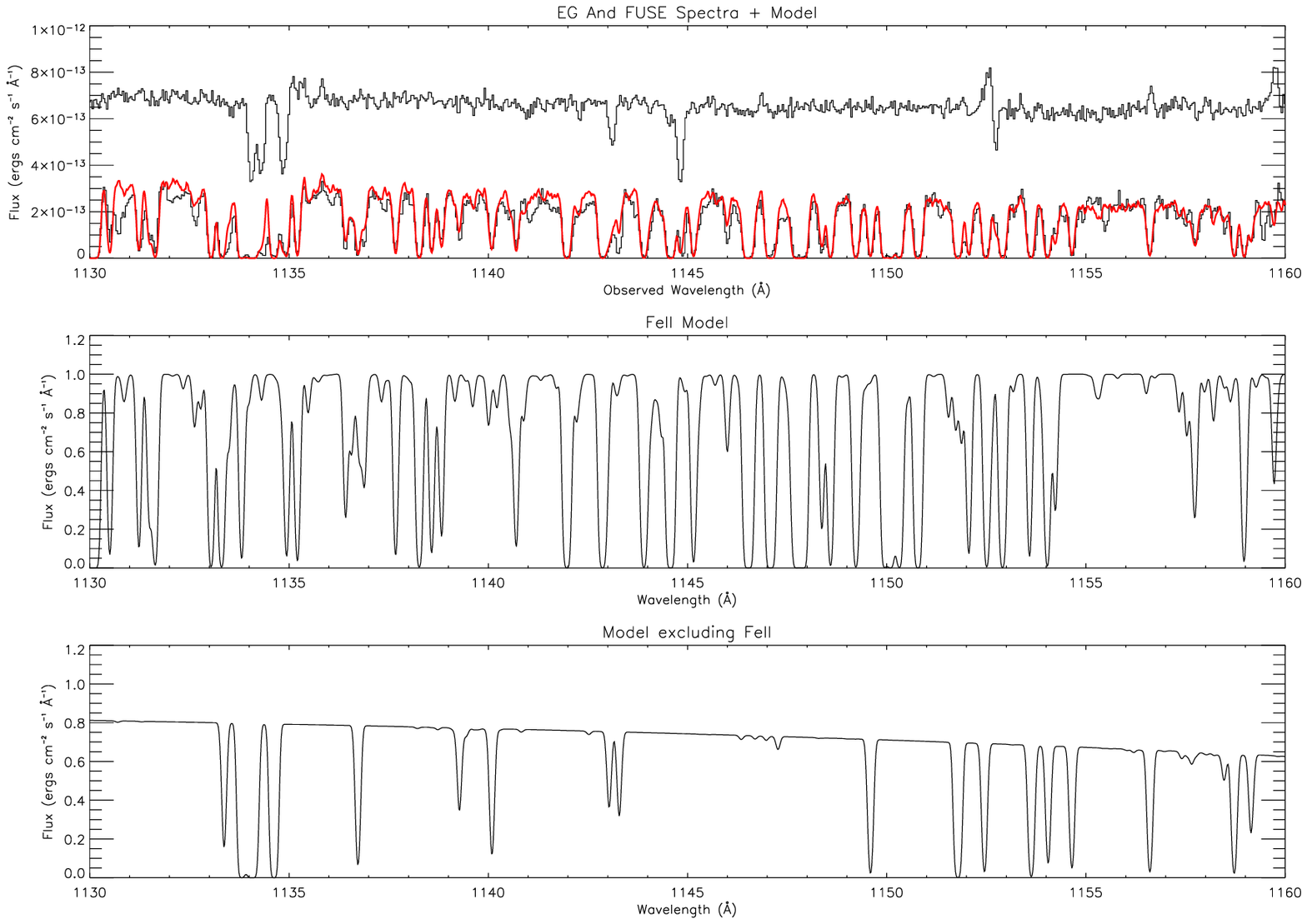 , width=18.3cm}
  \end{center}
\caption{ The top panel shows two {\it{FUSE}} spectra of EG And. The uneclipsed spectrum was taken at quadrature and is an unobscured view of the dwarf where the absorption lines are all interstellar (flux is offset by +3x10$^{13}$ for clarity). The absorbed spectrum was taken part-way during eclipse (phi=0.10) and a fit is overplotted (light). The fit was generated by placing absorbing material in the line of sight to the uneclipsed spectrum. The second panel shows the FeII contribution to the fit while the bottom panel shows the HI contribution along with a number of other species (in this wavelength region mostly NI, PII and NiII). Note that the blue wing of Lyman $\alpha$ defines the continuum shape. The FeII model was generated using atomic data from Raassen \& Uylings \label{fig5} }
\end{figure*}

\subsubsection{Temperature and Ionisation}

\begin{figure}[ht]
  \begin{center}
    \epsfig{file=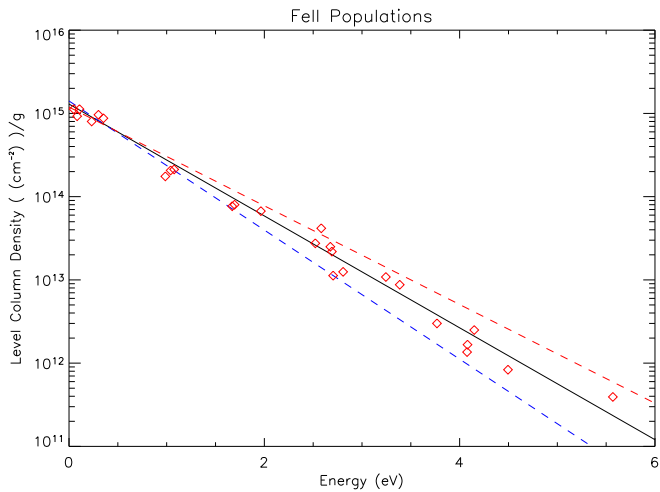 , width=9.3cm}
  \end{center}
\caption{By measuring the apparent absorbing column for a range of lines coming from different lower levels it is possible to examine the excitation structure in the gas. Here is an excitation diagram of FeII derived from one of the STIS spectra. The solid line corresponds to a total FeII column of 7x10$^{16}$ cm$^{-2}$ and a temperature of 7,500 K. The lower and higher dashed lines correspond to the same column but a temperature of 6,500 K and 8,500 K respectively. \label{fig6}}
\end{figure}

In the case of species such as FeII and NiII, which have a large amount of lines, it is apparent that there are absorption features arising from transitions from lower levels ranging from ground to levels $\sim$ 5eV above ground. This therefore makes it possible to calculate the level populations and examine the excitation structure. In all the absorbed spectra the FeII level populations can be explained by a simple collisional excitation model. By comparing the excitation diagrams with model ions populated collisionally it is possible to derive a temperature and total column density. We find that for the range of impact parameters observed ($\sim$ 2.2 - 3.7 R$_{RG}$) the wind temperature is consistently located within the range $\sim$ 6,000 - 8,000 K throughout eclipse (see Figure\ \ref{fig6}). 

This constancy in the conditions examined throughout eclipse is also apparent in the ionisation structure. The ionisation level stays constant throughout (for iron FeII is the only ionisation stage observed in absorption) until an impact parameter $\sim$3.7 is reached, after which there is swift onset of ionisation due to the radiation from the dwarf. This is most obvious for hydrogen where the large columns disappear rapidly. The variations in the columns of other species in the wind are observed to track that of hydrogen.

\subsubsection{Atomic Data}

Atomic data in the ultraviolet region of the spectrum (especially in the {\it{FUSE}} region) is notoriously uncertain for many transitions. Lines from lower levels with excited states up to $\sim$5 eV above ground are generally not observed in {\it{FUSE}} or STIS targets. The lines in this data originating from lower levels above ground can be used to check theoretical atomic oscillator strengths. For example the FeII lines in the {\it{FUSE}} data were fit using atomic data published by \cite{1995all..book.....K} and also data from \cite{1998JPhB...31.3137R} and it was found that the Raassen \& Uylings data provided a better match to the observed spectrum.

\subsection{Molecular Lines}
The {\it{FUSE}} wavelength range contains a large number of strong molecular hydrogen transitions from a range of energy levels. Indeed these spectra include a number of H$_2$ lines of interstellar origin. Our observations, therefore, are extremely sensitive to the presence of H$_2$ in the wind acceleration region at a distance  of $\sim$ 2.2 - 3.7 R$_{RG}$. No H$_2$ lines (or CO or H$_2$O lines in the STIS data) are detected, setting tight constraints on the molecular content of the wind at these distances above the limb. The upper limit of molecular hydrogen is found to be H$_2$/H $<$ 10$^{-8}$.

\section{Wind Acceleration}

By tracking the varying absorption profiles of neutral hydrogen at different phases it is possible to map the spatial extent and density of the wind. However, finding the run of the velocity profile using the HI column densities is a mathematically complicated inverse problem. \cite{1993A&A...274.1002K}  discuss this particular problem of reconstructing the density profile of a stellar wind in a binary system and have developed a technique for doing so. The equation to be inverted is:

\begin{equation}
N_H(b)=a \int_b^\infty \frac{dr}{\sqrt{r^2-b^2} \,  r \,v(r)} 
\end{equation}


where $b$ is the impact parameter of the two components, $a$ is a constant defined by the mass loss parameters of the system, $r$ is the distance from the giant and $v(r)$ is the velocity - ie. the variable we wish to determine.

As outlined by \cite{1993A&A...274.1002K} the continuity equation (1) is an integral equation of Fredholm's type but it can be brought to Abel's type by means of suitable transformations. Using this method it is then possible to invert the equation by an explicit diagonalization of the Abel's operator. The analysis assumes a neutral, spherical wind and a circular orbit. In practice the observed HI columns are parametrised by a polynomial and the velocity field is then obtained from the inverted Abel's operator. See \cite{1993A&A...274.1002K}  for a thorough discussion. This work implements the technique using some {\it{IUE}} EG And data and found that the wind acceleration region region lies at $\sim$ 2.5 R$_{RG}$ from the limb of the giant.  

\begin{figure*}[ht]
 \begin{center}
\includegraphics[scale=.43]{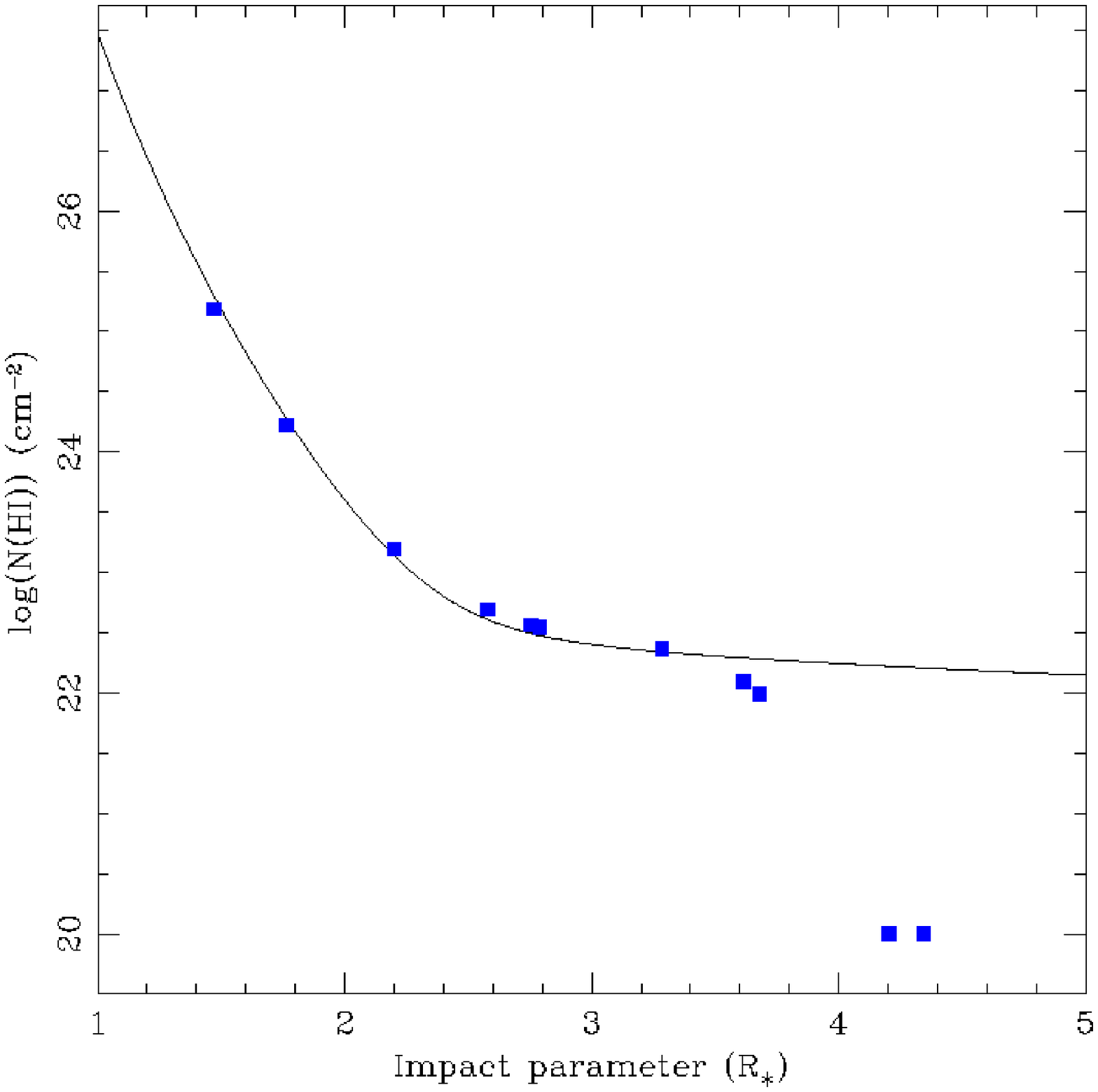}
\includegraphics[scale=.43]{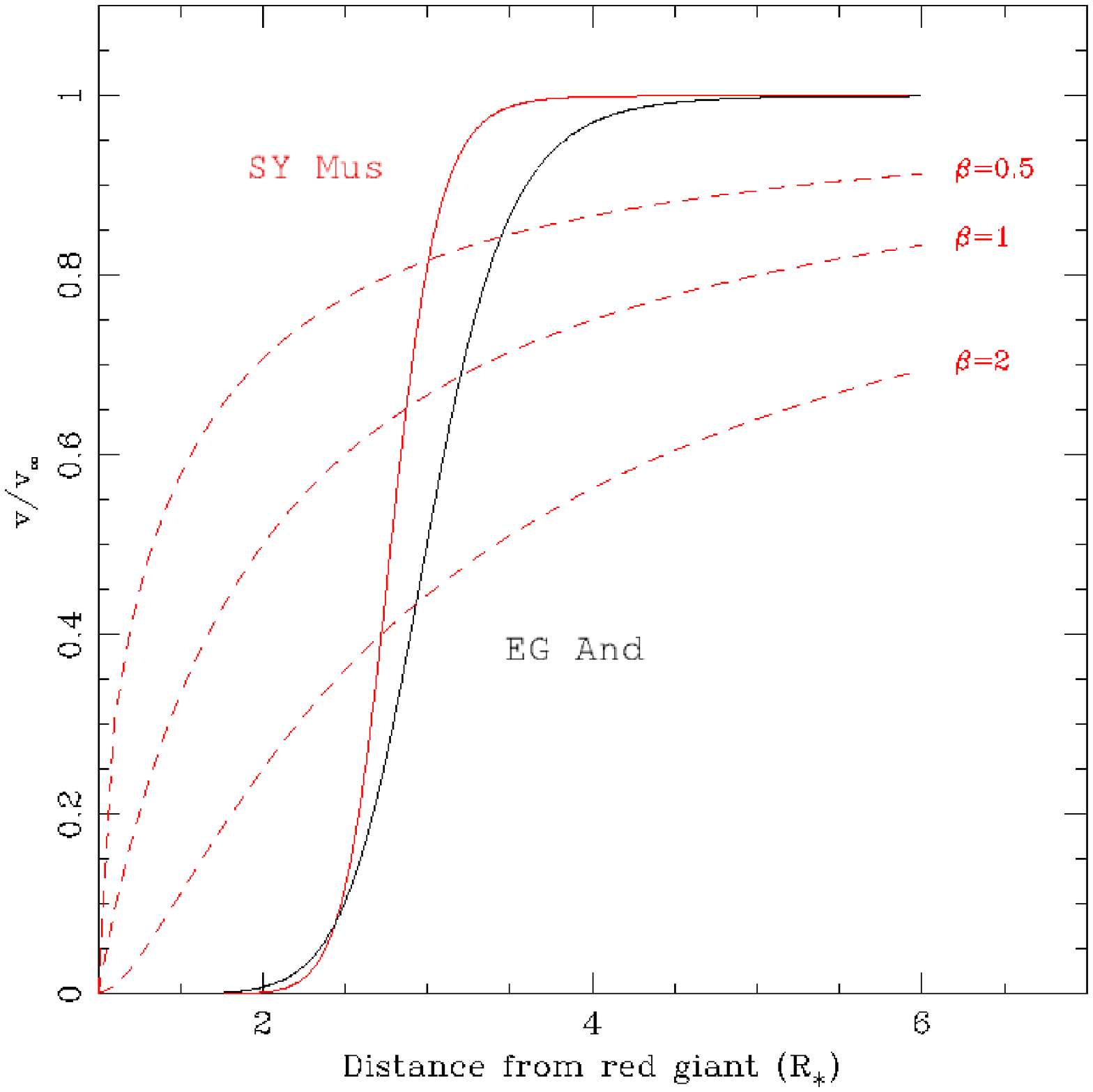}


 \end{center}
\caption{{\bf{Left:}} Model of the HI column density falloff as a function of distance from the giant primary in EG And. Overplotted are the HI columns derived from {\it{FUSE}} and STIS data. Note the deviation of the model from the datapoints at b $\approx$3.5R$_{RG}$ corresponding to the point where hydrogen is ionised by the WD. {\bf{Right:}} Plot of derived wind laws for the giants in \object{EG And} (black and SY Mus (grey) using the Abel inversion method. Note the deviation of the velocity profiles from the $\beta$ laws, commonly used to parameterise the wind profile in late giants.  \label{fig7}}
\end{figure*}

Plotted in Figure\ \ref{fig7} (left panel) are the HI column density datapoints measured on egress from the {\it{FUSE}} and STIS data plotted against impact parameter. Overplotted is a parametrised fit. The sudden drop in HI column at $b\sim$ $3.5 R_{RG}$ corresponds to the distance above the limb of the giant where its wind becomes ionised and the majority of hydrogen in the line of sight is no longer visible in the HI lines. It is apparent that the ionisation boundary is well-defined and the onset is swift.

Plotted in black in Figure\ \ref{fig7} (right panel) is the wind velocity profile obtained by inverting the model using the technique described above. Also shown are the $\beta$-laws which are commonly used to describe velocity profiles. These laws match observations for hot stars very well, however giant winds are known not to be radiatively driven and there is no good reason why the beta laws should successfully model these winds.  The disagreement with the wind in the cool giant is clear to see. 

Also plotted (in grey) is the velocity law derived for a similar systems, \object{SY Mus} by \cite{1999A&A...349..169D} where they obtained similar results using {\it{IUE}} data.

\section{Extending the Wind Analysis}

\begin{figure}[ht]
  \begin{center}
    \epsfig{file=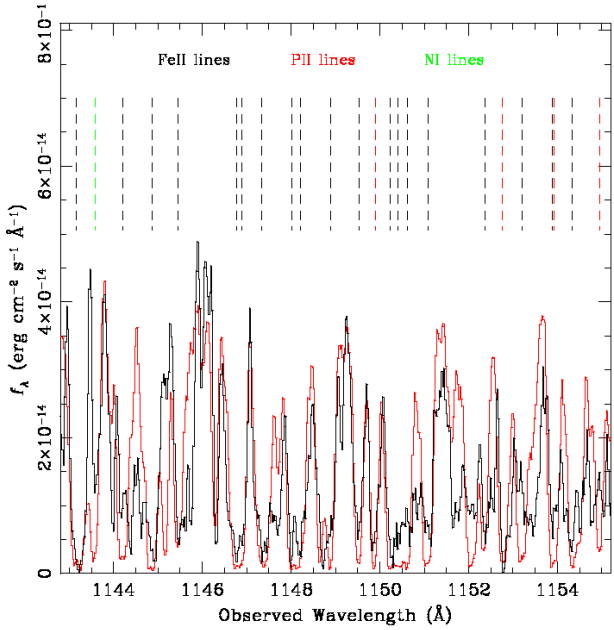 , width=9.cm}
  \end{center}
\caption{\textit{{\it{FUSE}}} spectrum of BF Cyg (black) plotted against an eclipsed spectrum of \object{EG And} (grey). Even though the phases are different it is apparent that the ionisation structure and composition of both winds are very similar. \label{fig8}}
\end{figure}

Our complete ultraviolet phase coverage of \object{EG And} (combined with our ground-based optical spectra and photometric monitoring) has allowed us to study the object in some detail. We have also been awarded {\it{FUSE}}  cycle 4 and cycle 5 time to obtain ephemeris observations of the similar eclipsing symbiotics, \object{BF Cyg} and \object{RW Hya} (see Appendix A for a {\it{FUSE}} observation log of \object{BF Cyg}). Our \object{BF Cyg} programme has recently finished we and are currently obtaining our \object{RW Hya} observations. Although in both cases the phase coverage is not as complete as it is for EG And, it is possible to use our understanding of the data to examine the similarities/differences in the winds in each system. For example a highly absorbed {\it{FUSE}} spectrum \object{BF Cyg} would be extremely difficult to interpret and model without the in-depth analysis of \object{EG And}. A simple comparison with a similarly attenuated \object{EG And} {\it{FUSE}} spectrum immediately shows that the ionisation level and composition of both winds are very similar (see Figure\ \ref{fig8}). Line identification and absorption line modelling has already been carried out for EG And allowing us to model the wind and velocity profile of \object{BF Cyg} quite successfully even allowing for the sparser coverage. 

It is hoped that extending the analysis to other objects will constrain the underlying wind driving mechanism at work in evolved giants.

\section{Discussion}

Symbiotic systems can be used as laboratories in which to perform detailed studies on giant winds in ways which would not be possible for isolated stars. Effects due to the presence of an ionising source close to the giants can be minimised by a careful target selection. In the case of \object{EG And} the dwarf is a factor $\sim$ 30 less luminous than the giant and the ionising effects closer to the giant appear minimal. In addition to wind analyses, much can be learned about the interaction of cool and hot gas in such systems, as well as providing constraints on atomic data in the far ultraviolet.

\subsection{Implications for Wind Driving Mechanisms}

\cite{1994AJ....108.1112V} present evidence for the lack of dust opacity in the wind of \object{EG And} - an argument based on extinction constraints from {\it{IUE}} data and far infrared fluxes from {\it{IRAS}}. Our {\it{FUSE}} and {\it{HST}} observations tighten the extinction constraints furthering the evidence that the wind cannot be dust driven. The authors proceed to discuss radiation pressure on molecular lines as a possible driver to accelerate the the material in the wind out to dust forming distances (also see \cite{1989ApJ...341L..95E}). However, our {\it{FUSE}} data provides tight upper limits on the presence of molecular hydrogen throughout the wind acceleration region thus ruling out any molecular line opacity at these impact parameters. 

Polarisation measurements have been carried out on \object{EG And} to determine whether a magnetic field is present (\cite{1990A&AS...86..227}). The results however, are inconclusive, and the small observed polarisation ($\sim$0.2\%-0.4\%) could be due to either scattering, a magnetic field or some other effect. As \cite{1994AJ....108.1112V} discuss, although high mass-loss rates and a large magnetic field are required for Alfven waves to be an effective wind driver, the observations don't rule this out and it is possible that magnetic fields play an important role. 

The composition and velocity profiles of the giant winds in \object{SY Mus} and \object{BF Cyg} are consistent with what is observed for \object{EG And}. The giants in these systems all have different mass-loss rates and spectral types which would suggest a wind driving mechanism that can be applied to non-dusty, late type giants generally. Further analysis of our \object{BF Cyg} and \object{RW Hya} {\it{FUSE}} data will provide further clues.

\subsection{Future Work}

Immediate future work concerns fully modelling the complicated phase-dependent absorption spectra of all three objects and tightening the constraints on the wind conditions. Work is also progressing on a wind model where the physical parameters are inputted and theoretical column densities are derived for different lines of sight. This allows us to fully disentangle the effects of the ionisation and mechanical disturbance of the dwarf on the giant wind. Preliminary results are consistent those derived from the Abel inversion method. 


\begin{acknowledgements}

This work is supported by Enterprise Ireland Basic Research grant SC/2002/370 from EU funded NDP

\end{acknowledgements}

\appendix

\section{UV Observations}

\begin{table}[!bht]
  \caption{\object{EG And} {\it{FUSE}} observations}
  \label{tab:table}
  \begin{center}
    \leavevmode
    \footnotesize
    \begin{tabular}[h]{cccc}
      \hline \\[-5pt]
      Obs. no. & Date      &  UV Phase      & Reduced Phase \\
      \hline \\[-5pt]
1 & 5 Jan 2000   & 1.7937 & -0.2063 \\
2 & 6 Aug 2000  & 2.2383 & 0.2383 \\
3 & 24 Nov 2000 &  2.4655 & 0.4655 \\
4 & 3 Sep 2001 &   3.0523 & 0.0523 \\
5 & 14 Sep 2001 &  3.0741 & 0.0741  \\
6 & 28 Sep 2001 &  3.1037 & 0.1037 \\
7 & 23 Oct 2001 & 3.1566 & 0.1566 \\
8 & 23 Aug 2002 &  3.7861 & -0.2139 \\
9 & 20 Oct 2002 &  3.9062 & -0.0938 \\
10 & 22 Oct 2002 &  3.9103 & -0.0897 \\
11 & 23 Oct 2002 &  3.9128 & -0.0872 \\
12 & 22 Jan 2003 &  4.1007 & 0.1007 \\
13 & 01 Dec 2003 &  4.7484 & -0.2516 \\

      \hline \\
      \end{tabular}
  \end{center}
\end{table}

\begin{table}[!bht]
  \caption{\object{EG And} STIS observations}
  \label{tab:table}
  \begin{center}
    \leavevmode
    \footnotesize
    \begin{tabular}[h]{cccc}
      \hline \\[-5pt]
      Obs. no. & Date      &  UV Phase      & Reduced Phase \\
      \hline \\[-5pt]

1 & 28 Aug 2002 & 3.796  & -0.204\\
2 & 16 Oct 2002 &  3.898  & -0.102\\
3 & 22 Dec 2002 &  4.036  & 0.036\\
4 & 18 Jan 2003 &  4.093 & 0.093\\
5 & 06 Feb 2003 &  4.131 & 0.131\\
6 & 16 Feb 2003 & 4.152 & 0.152\\
7 & 31 Jul 2003 &  4.495 & 0.495\\

      \hline \\
      \end{tabular}
  \end{center}
\end{table}

\begin{table}[!bht]
  \caption{BF Cyg {\it{FUSE}} Observations}
  \label{tab:table}
  \begin{center}
    \leavevmode
    \footnotesize
    \begin{tabular}[h]{cccc}
      \hline \\[-5pt]
      Obs. no. & Date      &  UV Phase      & Reduced Phase \\
      \hline \\[-5pt]

1 & 9 Aug 2000&  0.4895  & 0.4895    \\
2 & 6 May 2003&  1.8102  & -0.1898   \\
3 & 21 May 2003& 1.8296  & -0.1704   \\
4 & 4 Jun 2003&  3.8489  & -0.1511    \\ 
5 & 17 Jul 2003& 1.9053  & -0.0947    \\ 
6 & 30 Sep 2003& 2.0047  & 0.0047   \\ 
7 & 14 Apr 2004& 2.2649  & 0.2649 \\

      \hline \\
      \end{tabular}
  \end{center}
\end{table}

\end{document}